\documentclass[12pt]{article}
\usepackage{amsfonts}
\usepackage{bbm}
\usepackage{graphicx}
\usepackage{pgfplots}
\usepackage{booktabs}
\usepackage{subfigure}
\usepackage{mathrsfs}
\usepackage{color}
\usepackage{verbatim}
\usepackage{cancel}
\usepackage{geometry}             
\usepackage{amssymb}
\usepackage{amsmath}
\usepackage{epstopdf}
\usepackage{cases}
\usepackage{cite}
\usepackage{setspace}
\usepackage{array}
\usepackage[title]{appendix}
\usepackage{mathtools}
\usepackage{xcolor}
\newtagform{rednum}[\textcolor{red}]{(}{)}

\usepackage[hyperindex=true,
pdfstartview=FitH,
bookmarksnumbered=true,
bookmarksopen=true,
citecolor=blue,
linkcolor=blue,
colorlinks=true,
urlcolor=rossoCP3,
unicode]{hyperref}

\definecolor{rossoCP3}{cmyk}{0,.88,.77,.40}

\begin{document}

\title{\bf
Black hole thermodynamics from Iyer-Wald formalism with variable Newton's constant}
{\author{\small Xuan-Rui Chen${}^{1}$, Bin Wu${}^{1,2,3,4}$\thanks{{\em email}: \href{mailto:binwu@nwu.edu.cn}{binwu@nwu.edu.cn}}{ }, Zhen-Ming Xu${}^{1,2,3,4}$
        \vspace{5pt}\\
        \small $^{1}${\it School of Physics, Northwest University, Xi'an 710127, China}\\
        \small $^{2}${\it Shaanxi Key Laboratory for Theoretical Physics Frontiers, Xi'an 710127, China}\\
        \small $^{3}${\it Peng Huanwu Center for Fundamental Theory, Xi'an 710127, China}\\
	\small $^{4}${\it Fundamental Discipline Research Center for Quantum Science and technology of }\\
	\small {\it Shaanxi Province, Xi’an 710127, China}
        }
	
\date{}
\maketitle
\begin{spacing}{1.15}
\begin{abstract}
Recent advances in black hole thermodynamics have intensified efforts to investigate its thermodynamic framework and analogues to that of classical thermal systems. Departing from the extended phase space approach, which involves a variable cosmological constant, the restricted phase space formalism focusing on the variations of Newton's gravitational constant in black hole thermodynamics is proposed. This approach enables the construction of a self-consistent thermodynamic structure that features both the first law and the Euler relation. In this paper, we demonstrate that for any diffeomorphism-invariant gravitational theory, the Iyer-Wald formalism is applicable to restricted phase space thermodynamics and construct a self-consistent extensive black hole thermodynamic system. Our work thereby establishes the Iyer-Wald approach as the geometric foundation for black hole thermodynamics with variable Newton's constant.
\end{abstract}
\section{Introduction}
As a universal theory, thermodynamics is widely employed in various condensed matter systems, celestial bodies, quantum systems, and even non-extensive systems such as those with small scales or long-range interactions \cite{tsallis2003,ABULMAGD2007450,e17010052,Nivanen_2003Kaniadakis,Nojiri:2023ikl}. Based on the universality of thermodynamics, it is natural to apply it to black hole systems to analyze their thermal properties. Intriguingly, it is found that the AdS charged black hole exhibits a phase transition analogous to the liquid-gas phase transition in the Van der Waals system \cite{Chamblin:1999tk,Chamblin:1999hg,Kubiznak:2012wp}. Furthermore, when the negative cosmological constant $\Lambda$ is interpreted as thermal pressure, it has been discovered that black holes exhibit rich thermodynamic behaviors \cite{Kastor:2009wy,Dolan:2010ha,Dolan:2011xt,Cvetic:2010jb}. This framework is called extended 
phase space (EPS) black hole thermodynamics, and is now also known as black hole chemistry \cite{Frassino:2015oca,Kubiznak:2016qmn}. Reference \cite{Mann:2025xrb} provides a comprehensive review of 
black hole chemistry and its recent developments.

With the development of the AdS/CFT correspondence\cite{Gubser:1998bc,Witten:1998qj,Maldacena:1997re}, it is conjectured that the thermodynamic properties of AdS black holes can be understood from a holographic perspective\cite{Kastor:2014dra,Zhang:2014uoa,Karch:2015rpa}. It is suggested that the variation of the cosmological constant in the gravitational theory corresponds to the variation of the number of colors $N$ or the degrees of freedom $N^2$ in the boundary theory. This observation has led to the development of holographic thermodynamics, which restores an Euler-like relation among the thermodynamic quantities of the dual CFT \cite{Visser:2021eqk,Cong:2021fnf,Cong:2021jgb,Ahmed:2023dnh,Ahmed:2023snm,Yang:2024krx}.

Nevertheless, the Smarr relation in the EPS framework takes a non-homogeneous form. While this relation can be derived from dimensional analysis and, more fundamentally, from the Iyer-Wald formalism, the physical interpretation of its coefficients is not as directly transparent as in the Euler relation of classical thermodynamics. In the latter, all coefficients are equal to unity, admitting a clear interpretation in terms of the equal contribution of each extensive variable to the total energy. This raises the question of whether black hole thermodynamics can be formulated in a way that more directly incorporates this extensivity structure.

To address this, the restricted phase space (RPS) approach has been developed, where the cosmological constant $\Lambda$ is kept fixed while Newton's constant $G$ is allowed to vary \cite{Zeyuan:2021uol,Zhao:2022dgc,Gao:2021xtt}. Through the holographic dictionary $C \propto l^{d-2}/G$ \cite{Henningson:1998gx,Freedman:1999gp,Myers:2010xs,Myers:2010tj}, the central charge $C$ of the dual CFT becomes a genuine extensive thermodynamic variable, yielding the homogeneous Euler relation, which matches the extensivity structure of classical thermodynamics. Note that a change in $C$ can arise either from a change in $l$ at fixed $G$ (the EPS approach), or from a change in $G$ at fixed $l$ (the RPS approach). In this work we adopt the latter, so the variation of $C$ is entirely determined by that of $G$. The variation of $G$ is thus a means to achieve extensivity, and it is the physical operation in the bulk that corresponds, through holography, to varying the central charge $C$ in the boundary CFT.

The possibility of varying $G$ is motivated by a variety of physical settings. In string theory and its low-energy effective theories, many seemingly fundamental coupling constants are not fixed parameters, but are controlled by the vacuum expectation values of scalar fields such as the dilaton, leading to spacetime-dependent variations of $G$ \cite{Wu:1986ac,Damour:1994zq,Gasperini:2002bn,Uzan:2010pm}. From a cosmological perspective, such variations can influence the phenomenology of the dark sector and the late-time accelerated expansion of the Universe \cite{Pace:2013pea,Paliathanasis:2018fve,MV:2026ixo}. Moreover, recent work has shown that quantum-induced corrections to the generalized entropy can be interpreted as an effective shift in $G$ \cite{Lu:2024ppa,Wu:2026irp}. These considerations suggest that $G$ should be treated as an effective gravitational coupling rather than an immutable constant.

Despite these motivations, the existing restricted phase space framework lacks a first-principles explanation, and its geometric origins remain elusive. Previous methods for extracting thermodynamic quantities typically relied on defining the chemical potential through identities such as $\mu \equiv (\partial M/\partial C)_{S,J...}$, or extracting it indirectly from the Euclidean action, neither of which provides a geometric foundation. 

The Iyer-Wald formalism has proven to be a powerful geometric tool for deriving black hole thermodynamics from first principles \cite{Wald:1993nt,Iyer:1994ys}. It provides a Lagrangian-based framework for constructing Noether currents and charges associated with spacetime symmetries. The formalism has been extended to the EPS, where $\Lambda$ varies, and the Smarr relation becomes non-homogeneous \cite{Xiao:2023lap}. A general formalism has also been developed to treat coupling constants as conserved charges, but within the EPS framework with $G$ fixed \cite{Hajian:2023bhq}. In both cases, the resulting thermodynamic relations are non-homogeneous and do not align with the extensivity structure of classical thermodynamics. Furthermore, the extended Iyer-Wald formalism has also been successfully applied to numerous scenarios, including dyonic black holes in Einstein-Maxwell-Chern-Simons gravity \cite{Cai:2024tyv}, exact isohomogeneous transformation-based \cite{Campos:2025nnk}, maximally symmetric vacuum solution \cite{Tavlayan:2024zbl}, and generalized free energy landscapes \cite{Wu:2025jes}.

In this paper, we provide such a first-principles geometric foundation for the RPS thermodynamics by extending the Iyer-Wald formalism to incorporate variations of $G$ in a way that is conceptually distinct from previous EPS-based extensions. We systematically derive the RPS first law and the homogeneous Euler relation directly from the Lagrangian and Noether charges, demonstrating that the extensivity of RPS thermodynamics is not an ad hoc input but emerges naturally from the geometry. We verify the consistency of our framework through 
applications to three canonical black hole solutions, Schwarzschild-AdS, Reissner-Nordström-AdS, and Kerr-AdS, and cross-check our geometric results with independent Euclidean action calculations.

The organization of this paper is as follows. In Section \ref{II}, we review the standard Iyer-Wald formalism. In Section \ref{III}, we extend the formalism to incorporate variations of $G$ and apply it to the three black hole solutions. Finally, our conclusions are presented in Section \ref{IV}.

\section{Iyer-Wald formalism for conventional black hole thermodynamics}\label{II}
To establish the necessary theoretical background for our subsequent extension, we first provide a concise review of the standard geometric Iyer-Wald formalism \cite{Wald:1993nt,Iyer:1994ys}. The Iyer-Wald formalism starts from the Lagrangian density of general relativity. It is well established that for any diffeomorphism-invariant gravitational theory, the conventional laws of black hole thermodynamics can be readily derived using the Iyer-Wald formalism.

Let us consider a Lagrangian $\mathbf{L}$ expressed as $\mathbf{L}=L \boldsymbol{\epsilon}$ that depends solely on the dynamical fields $\phi\equiv \{g_{\mu \nu},\psi\}$. Here, $\boldsymbol{\epsilon}$ is the volume form of the $D=d+1$ dimensional space-time, $g_{\mu \nu}$ represents the metric tensor, and $\psi$ denotes other matter fields. We focus on diffeomorphism-invariant theories, for any diffeomorphism $\Phi: M \rightarrow M$, the Lagrangian $\mathbf{L}$ satisfies the condition
\begin{align}
	\mathbf{L}[\Phi^{\ast} \phi]=\Phi^{\ast}\mathbf{L}[\phi],
\end{align}
here, $\Phi^{\ast}$ denotes the pullback by the diffeomorphism $\Phi$, which maps the Lagrangian 
$\mathbf{L}$ from the manifold $M$ to itself while preserving its functional form with respect to the fields $\phi$. The variation of the Lagrangian can be written as
\begin{align}
	\delta \mathbf{L}=\mathbf{E}^{\phi}\delta  \phi+ d \boldsymbol{\Theta}[\delta \phi], \label{1}
\end{align}
where $\mathbf{E}^{\phi}$ represents the equations of motion derived from the Lagrangian, and
$\boldsymbol{\Theta}[\delta \phi]$ is a total derivative term. For any smooth vector field $\xi$, the corresponding variation of the dynamical fields is given by the Lie derivative, $\delta_\xi \phi = \mathcal{L}_\xi \phi$. Similarly, the variation of the Lagrangian is $\delta_\xi \mathbf{L} = \mathcal{L}_\xi \mathbf{L}$. 
	
In differential geometry, the Lie derivative acting on a differential form can be expanded using Cartan's formula $\mathcal{L}_\xi \mathbf{L} = \xi \cdot d\mathbf{L} + d(\xi \cdot \mathbf{L})$. Since $\mathbf{L}$ is already a top-dimensional form on the spacetime manifold, its exterior derivative must vanish (i.e., $d\mathbf{L} = 0$). Therefore, the geometric variation of the Lagrangian can be completely simplified to an exact differential form
\begin{align}
	\delta_\xi \mathbf{L}= \mathcal{L}_\xi \mathbf{L} = d(\xi \cdot \mathbf{L}).
\end{align}
Combining this with the variation of the Lagrangian in Eq.\eqref{1}, we obtain
\begin{align}
	d(\xi \cdot \mathbf{L})=\delta_\xi \mathbf{L}=\mathbf{E}^{\phi}\delta_\xi \phi+ d \boldsymbol{\Theta}[\delta_\xi \phi]= \mathbf{E}^\phi \mathcal{L}_\xi \phi + d\boldsymbol{\Theta}[\mathcal{L}_\xi \phi].
\end{align}
Moving the terms containing the exterior derivative operator $d$ to one side of the equation, we obtain the following after simplification
\begin{align}
d( \boldsymbol{\Theta}[\mathcal{L}_\xi \phi] - \xi \cdot \mathbf{L} ) = -\mathbf{E}^\phi \mathcal{L}_\xi \phi.
\end{align}
From this, by observing the combination of terms inside the parentheses on the left-hand side of the above equation, the Noether current $\mathbf{J}_\xi$ is defined as follows \cite{Lee:1990nz}
\begin{align}
	\mathbf{J}_\xi \equiv \boldsymbol{\Theta}[\mathcal{L}_\xi \phi]-\xi  \cdot \mathbf{L},
	\label{NC}
\end{align}
under the on-shell condition $\mathbf{E}^{\phi}=0$, the Noether current is closed, $d\mathbf{J}_\xi = 0$. Thus it can be expressed as $\mathbf{J}_\xi = d\mathbf{Q}_\xi$, where $\mathbf{Q}_\xi$ is the Noether charge associated with $\xi$.

By taking the variation of the Noether current in Eq.(\ref{NC}), we can derive the key identity that is crucial for establishing the first law of the thermodynamics of black holes. By substituting the variation of the Lagrangian from Eq.(\ref{1}) and imposing the on-shell condition $\mathbf{E}^{\phi}=0$, we can get
\begin{align}
	\delta\mathbf{J}_\xi=\delta\boldsymbol{\Theta}[\mathcal{L}_\xi \phi]-\xi  \cdot \delta \mathbf{L}=
	\delta\boldsymbol{\Theta}[\mathcal{L}_\xi \phi]-\xi  \cdot d \boldsymbol{\Theta}[\delta \phi].\label{J1}
\end{align}
Applying the identity of the Lie derivative $\mathcal{L}_\xi \boldsymbol{\Theta}[\delta \phi]=\xi  \cdot d\boldsymbol{\Theta}[\delta \phi]+d(\xi  \cdot \boldsymbol{\Theta}[\delta \phi])$, the last term in the above expression can be replaced, resulting in
\begin{align}
	\delta \mathbf{J}_\xi=\delta\boldsymbol{\Theta}[\mathcal{L}_\xi \phi]-\mathcal{L}_\xi \boldsymbol{\Theta}[\delta \phi]+d(\xi  \cdot \boldsymbol{\Theta}[\delta \phi]).
\end{align}
In the covariant phase space method, the definition of the symplectic current is given by $\boldsymbol{\omega}[\delta_1\phi,\delta_2\phi] = \delta_1\boldsymbol{\Theta}[\delta_2 \phi] - \delta_2\boldsymbol{\Theta}[\delta_1 \phi]$. This definition dictates that the first two terms in the derived expression combine to form the symplectic current $\boldsymbol{\omega}[\delta \phi,\mathcal{L}_\xi \phi]=\delta\boldsymbol{\Theta}[\mathcal{L}_\xi \phi] - \mathcal{L}_\xi \boldsymbol{\Theta}[\delta \phi]$. From this, we obtain the simplified equation
\begin{align}
	\delta \mathbf{J}_\xi = \boldsymbol{\omega}[\delta \phi,\mathcal{L}_\xi \phi] + d(\xi \cdot \boldsymbol{\Theta}[\delta \phi]).\label{J2}
\end{align}
For a Killing vector $\xi$ characterized by the property $\mathcal{L}_\xi \phi=0$, the symplectic current naturally vanishes. By utilizing the property of the exact form $\delta \mathbf{J}_\xi = \delta d\mathbf{Q}_\xi = d\delta \mathbf{Q}_\xi$, the above equation reduces to
\begin{equation}
    d\delta \mathbf{Q}_\xi = d(\xi \cdot \boldsymbol{\Theta}[\delta \phi]).
\end{equation}
Rearranging this equation, we obtain the first differential formula
\begin{equation}
    d(\delta \mathbf{Q}_\xi - \xi \cdot \boldsymbol{\Theta}[\delta \phi]) = 0. \label{dd}
\end{equation}
In addition, from the on-shell identity $\mathbf{J}_\xi = d\mathbf{Q}_\xi$, and noting that $\mathcal{L}_\xi \phi = 0$,  we directly obtain the second formula from Eq.\eqref{NC}
\begin{equation} 
    d\mathbf{Q}_\xi = -\xi \cdot \mathbf{L}. \label{dQ}
\end{equation}
Moreover, by integrating the resulting expression Eqs.(\ref{dd}) and (\ref{dQ}) over a hypersurface $\Sigma$, we derive the foundational thermodynamic relations governing black hole thermodynamics.
\begin{align}
	\int_{\partial \Sigma} &\left(  \delta \mathbf{Q}_{\xi} -  \xi \cdot \boldsymbol{\Theta}[\delta \phi] \right)  =0, \label{intfor1} \\
	& \int_{\partial \Sigma} \mathbf{Q}_{\xi} = -\int_{\Sigma}\xi  \cdot \mathbf{L}. \label{intfor2}
\end{align}
The horizon $S_H$ functions as an intrinsic integration boundary where the Killing vector field $\xi$ vanishes, namely $\xi_H=0$. The other integration boundary may take the form of any surface $S_r$ with an arbitrary radius surrounding the black hole. Later, we will notice that the specific radial value $r$ of the surface, irrespective of its location, has no bearing on the results. In particular, within the framework of Wald entropy, it can be shown that the relations $ \int_{S_H} \delta \mathbf{Q}_{\xi_H}=T\delta S$ and $\int_{S_H} \mathbf{Q}_{\xi_H} = TS$ hold \cite{Wald:1993nt,Iyer:1994ys}. Consequently, Eq.\eqref{intfor1} leads to the first law of black hole thermodynamics, while Eq.\eqref{intfor2} yields the corresponding integrated relation.

\section{Iyer-Wald formalism with variable Newton's constant}\label{III}
The primary objective of this work is to establish a geometrical foundation for the thermodynamic formulation of black holes in the restricted phase space based on the Iyer-Wald formalism. Unlike the conventional variational framework of thermodynamics, the restricted phase space treats Newton's gravitational constant $G$ as a variable parameter, which necessitates an appropriate extension of the standard variational procedure.

To incorporate the variation of Newton's constant \(G\), we introduce a generalized variation operator \(\tilde{\delta}\), which acts on the metric as
\begin{equation}
\tilde{\delta} g_{\mu \nu} = \delta g_{\mu \nu} + \frac{\partial g_{\mu \nu}}{\partial G} \tilde{\delta} G.
\end{equation}
Applying \(\tilde{\delta}\) to the Lagrangian \(\mathbf{L}(g_{\mu\nu}, G)\), the total variation splits into two contributions, one from varying the metric at fixed \(G\), and another from the variation of \(G\), which includes the explicit dependence of \(\mathbf{L}\) on \(G\) and the implicit dependence through \(g_{\mu\nu}\). Combined, the metric-dependent part can be written compactly as \((\partial \mathbf{L}/\partial g_{\mu\nu})\tilde{\delta} g_{\mu\nu}\), yielding the complete variation
\begin{align}
	\tilde{\delta}\mathbf{L}
	=
	\frac{\partial \mathbf{L}}{\partial g_{\mu\nu}}\tilde{\delta} g_{\mu\nu}
	+
	\frac{\partial \mathbf{L}}{\partial G}\tilde{\delta}G.
\end{align}	
Next, we apply the standard variational identity of the Iyer-Wald formalism to the first term. This transformation simplifies the variation of the Lagrangian into
\begin{align}
	\begin{split}
		\tilde{\delta} \mathbf{L}
		=\mathbf{E}^{\phi} \tilde{\delta}  \phi+  d \boldsymbol{\Theta}[\tilde{\delta} \phi]+F_g\tilde{\delta}G\boldsymbol{\epsilon},\label{deltaL}
	\end{split}
\end{align}
where $F_g=(\partial \mathbf{L}/\partial G)_{g_{\mu\nu}}$ and $\boldsymbol{\epsilon}$ denotes the volume form. Here, $\mathbf{E}^{\phi}$ represents the equations of motion derived from the Lagrangian, $\tilde{\delta} \phi$ denotes the variation of the fields, and $\boldsymbol{\Theta}[\tilde{\delta}\phi]$ is a total derivative term. 

Subsequently, we rederive Eq.\eqref{intfor1} by utilizing the Killing vector $\xi_H$ at the horizon. Specifically, based on the definition of the Noether current in Eq.\eqref{NC}, substituting the variation of the Lagrangian in Eq.\eqref{deltaL} into the variation of the generalized current and taking into account the on-shell condition $\mathbf{E}^{\phi}=0$, we can obtain the following relationship
\begin{align}
		\tilde{\delta} \mathbf{J}_{\xi_H} =\tilde{\delta}\boldsymbol{\Theta}[\mathcal{L}_{\xi_H} \phi]-\xi_H \cdot \tilde{\delta}\mathbf{L}
		=\tilde{\delta}\boldsymbol{\Theta}[\mathcal{L}_{\xi_H} \phi] - \xi_H \cdot d\boldsymbol{\Theta}[\tilde{\delta}\phi] - F_g \xi_H \cdot \tilde{\delta}G\boldsymbol{\epsilon}.\label{J3}
\end{align}
Following the calculation logic between Eqs.\eqref{J1} and \eqref{dQ}, we use the Lie derivative to handle the $\xi_H \cdot d\boldsymbol{\Theta}[\tilde{\delta}\phi]$ term. Incorporating the definition of the symplectic current $\boldsymbol{\omega}[\tilde{\delta}\phi, \mathcal{L}_{\xi_H} \phi]$, the equation simplifies to
\begin{equation}
	\begin{aligned}
		\tilde{\delta} \mathbf{J}_{\xi_H} 
		&=\tilde{\delta}\boldsymbol{\Theta}[\mathcal{L}_{\xi_H} \phi] - \mathcal{L}_{\xi_H} \boldsymbol{\Theta}[\tilde{\delta}\phi] + d(\xi_H \cdot \boldsymbol{\Theta}[\tilde{\delta}\phi])- F_g \xi_H \cdot \tilde{\delta}G\boldsymbol{\epsilon}\\
		&=\boldsymbol{\omega}[\tilde{\delta}\phi, \mathcal{L}_{\xi_H} \phi]+ d(\xi_H \cdot \boldsymbol{\Theta}[\tilde{\delta}\phi])- F_g \xi_H \cdot \tilde{\delta}G\boldsymbol{\epsilon}\\
		&=d(\xi_H \cdot \boldsymbol{\Theta}[\tilde{\delta}\phi])- F_g \xi_H \cdot \tilde{\delta}G\boldsymbol{\epsilon},\label{JJ}
	\end{aligned}
	\end{equation}
where the symplectic current vanishes because $\mathcal{L}_{\xi_H} \phi = 0$ for the horizon Killing vector. Using $\tilde{\delta}\mathbf{J}_{\xi_H} = \tilde{\delta}(d\mathbf{Q}_{\xi_H}) = d(\tilde{\delta}\mathbf{Q}_{\xi_H})$ and rearranging the terms, we can get
\begin{align}
	d( \tilde{\delta} \mathbf{Q}_{\xi_H} - \xi_H \cdot \boldsymbol{\Theta}[\tilde{\delta} \phi]) = -F_g\xi_H\cdot\tilde{\delta}G\boldsymbol{\epsilon}.\label{L4}
\end{align}
Furthermore, combining the on-shell identity $\mathbf{J}_{\xi_H} = d\mathbf{Q}_{\xi_H}$ with the definition of the Noether current, we directly obtain
\begin{align}
d \mathbf{Q}_{\xi_H} = -\xi_H \cdot \mathbf{L}.\label{L5}
\end{align}
Subsequently, integrating Eqs.(\ref{L4}) and (\ref{L5}) over a hypersurface $\Sigma$ bounded by the event horizon $S_H$ and an arbitrary spatial surface $S_r$, we arrive at
\begin{align}
	\int_{S_r}\left(\tilde{\delta} \mathbf{Q}_{\xi_H} - {\xi_H}\cdot \boldsymbol{\Theta}[\tilde{\delta} \phi] \right) - \int_{S_H} \tilde{\delta} \mathbf{Q}_{\xi_H}&=-\tilde{\delta}G \int_{\Sigma}F_g{\xi_H}\cdot\boldsymbol{\epsilon}, \label{SAdS} \\
	\int_{S_r} \mathbf{Q}_{\xi_H}-\int_{S_H} \mathbf{Q}_{\xi_H}=-\int_{\Sigma}&\xi_H \cdot \mathbf{L}. \label{Euler}
\end{align}
From this, we have obtained the first law of thermodynamics of black holes with a variable Newton's constant and its corresponding integral relationship.

In the restricted phase space, $l$ is fixed, so the variation of $C$ is entirely determined by that of $G$. This is conceptually distinct from the EPS approach, where $G$ is fixed, and the variation of $C$ would arise from changes in $l$. Thus, in what follows, we express our results in terms of $\tilde{\delta}C$, while the underlying bulk variation is always that of $G$.

Before turning to specific solutions, we outline the general procedure by which Eq.\eqref{SAdS} will be applied. For a given black hole solution, the steps are as follows:

(1) From the action, read off the symplectic potential $\boldsymbol{\Theta}$ and the Noether charge $\mathbf{Q}_\xi$ associated with the horizon Killing vector $\xi_H$.

(2) Substitute these into Eq.\eqref{SAdS} and evaluate the boundary terms at $S_H$ (where $\xi_H=0$) and at $S_\infty$. The horizon term yields $T\tilde{\delta} S$, while the infinity term yields $\tilde{\delta} M$ and any work terms.

(3) Convert the $\tilde{\delta} G$ term to $\tilde{\delta} C$ via $C = l^2/G$, and read off the chemical potential $\mu$. The result is the first law; the integrated form gives the Euler relation.

In the following three subsections, we apply this procedure to Schwarzschild-AdS, Reissner-Nordström-AdS, and Kerr-AdS black holes. The three cases form a progressive sequence from neutral to charged, and from static to rotating, illustrating how the formalism accommodates additional conserved charges while maintaining the same underlying geometric structure. In the following subsections, we switch from the abstract differential-form notation to explicit tensor-component notation for practical calculations. Accordingly, the boldface symbols will be replaced by their component forms $L$, $\Theta^\mu$, $J_\xi^\mu$, and $Q_\xi^{\mu\nu}$.

\subsection{Schwarzschild-AdS black hole}
We start with the simplest model of a Schwarzschild-AdS (SAdS) black hole. As the most fundamental and classical physical model, it contains only the basic gravitational degrees of freedom. It serves to demonstrate the core construction process of the Lagrangian current in the extended Iyer-Wald formalism with a variable $G$. The bulk action is given by
\begin{align}
        I=\frac{1}{16\pi G}\int_{\mathcal{M}}d^4x \sqrt{-g}(R-2\Lambda),
\end{align}
with the negative cosmological constant $\Lambda=-3/l^2$ related to the AdS radius $l$ and Newton's constant $G$. The metric of the SAdS black hole takes the form
\begin{align}
        d s^{2}&=-f(r) d t^{2}+\frac{1}{f(r)} d r^{2}+r^{2}(d\theta^{2}+\sin^2{\theta}d\phi^{2}),
\end{align}
with the metric function
\begin{align}
        f(r)=1-\frac{2GM}{r}+\frac{r^2}{l^2}.
\end{align}
Here, $M$ is the mass of the black hole, which can be obtained by solving $f(r_{h})=0$, yielding
\begin{align}
	M = \frac{r_h}{2G} \left( 1 + \frac{r_h^2}{l^2} \right),\label{M}
\end{align}
where $r_h$ is the radius of the event horizon. The associated thermodynamic quantities, namely the Hawking temperature and the Bekenstein-Hawking entropy, evaluate straightforwardly to
\begin{equation}
	T = \frac{f'(r_h)}{4\pi} = \frac{1}{4\pi r_h} \left( 1 + \frac{3r_h^2}{l^2} \right), \quad S = \frac{\pi r_h^2}{G}.
\end{equation}
These fundamental quantities establish the foundation for verifying the geometric derivation of the restricted phase space thermodynamics.

Following the standard textbook approach, the complete variation of the action with respect to the metric tensor yields
	\begin{equation}
		\begin{aligned}
			\delta I &= \frac{1}{16\pi G}\int_{\mathcal{M}}d^{4}x \left[ \delta(\sqrt{-g})(R-2\Lambda) + \sqrt{-g}\delta (R-2\Lambda) \right] \\
			&= \frac{1}{16\pi G}\int_{\mathcal{M}}d^{4}x \sqrt{-g} \Bigg[ \left( -R^{\mu\nu} + \frac{1}{2}g^{\mu\nu}R - \Lambda g^{\mu\nu} \right)\delta g_{\mu\nu} \\
			&\quad + \nabla_{\mu} \left( g^{\mu \alpha} \nabla^\nu \delta g_{\alpha \nu} - g^{\alpha\beta}\nabla^{\mu}\delta g_{\alpha\beta} \right) \Bigg].
		\end{aligned}
\end{equation}
Among them, the first term in the integrand corresponds to the equation of motion of the gravitational field, and the total derivative term can be extracted from the second term of the above equation, yielding
\begin{align}
        \Theta^\mu (\delta g_{\mu \nu})\equiv \frac{1}{16 \pi G} (g^{\mu \alpha} \nabla^\nu \delta g_{\alpha \nu} - g^{ \alpha \beta} \nabla^{\mu} \delta g_{ \alpha \beta} ). \label{einsteintheta}
\end{align}
Under a diffeomorphism generated by the vector field $\xi$, the variation of the metric tensor is given by the Lie derivative $\delta g_{\mu\nu} = \mathcal{L}_\xi g_{\mu\nu} = \nabla_\mu \xi_\nu + \nabla_\nu \xi_\mu$. Substituting this into the symplectic potential and contracting the indices simplifies the expression to
\begin{equation}
	\Theta^\mu(\mathcal{L}_\xi g_{\mu \nu})= \frac{1}{16 \pi G} \left( \Box \xi^\mu + \nabla^\nu \nabla^\mu \xi_\nu - 2\nabla^\mu \nabla_\nu \xi^\nu \right),
\end{equation}
where $\Box \xi^\mu \equiv \nabla^\nu \nabla_\nu \xi^\mu$. To cast the derivative terms into a purely antisymmetric form, we apply the Ricci identity $[\nabla^\nu, \nabla^\mu]\xi_\nu = R^{\mu\nu}\xi_\nu$. This reorganizes the symplectic potential into a total divergence accompanied by a curvature term, as follows
\begin{equation}
	\Theta^\mu(\mathcal{L}_\xi g_{\mu \nu}) = \frac{1}{16 \pi G} \left[ -\nabla_\nu (\nabla^\mu \xi^\nu - \nabla^\nu \xi^\mu) + 2R^{\mu\nu}\xi_\nu \right].
\end{equation}
Based on the definition of the Noether current $J_\xi^\mu = \Theta^\mu(\mathcal{L}_\xi g_{\mu \nu}) - \xi^\mu L$, and substituting the Lagrangian $L = \frac{1}{16\pi G}(R - 2\Lambda)$ along with the evaluated symplectic potential, $J_\xi^\mu$ can be decomposed into an exact divergence and a term proportional to the equations of motion, yielding
\begin{equation}
	\begin{aligned}
		J^\mu_\xi &= \nabla_\nu \left[ -\frac{1}{16\pi G}(\nabla^\mu \xi^\nu - \nabla^\nu \xi^\mu) \right] + 2 \left[ \frac{1}{16\pi G} \left( R^{\mu\nu} - \frac{1}{2}g^{\mu\nu}R + \Lambda g^{\mu\nu} \right) \right] \xi_\nu\\ 
		&= \nabla_\nu Q_\xi^{\mu\nu} + 2 E^{\mu\nu}\xi_\nu. 
	\end{aligned}
\end{equation}
Imposing the on-shell condition $E^{\mu\nu} = 0$, $J_\xi^\mu$ strictly reduces to the divergence of the antisymmetric Noether charge tensor \cite{Iyer:1994ys}
\begin{equation}
	Q_\xi^{\mu\nu} \equiv -\frac{1}{16\pi G}  \left (  \nabla^\mu \xi^\nu -\nabla^{\nu}\xi^\mu  \right ). \label{einq}
\end{equation}	
Here Newton's constant $G$ should be regarded as a variable with respect to the symbol $\tilde{\delta}$. For SAdS black holes, the horizon
Killing vector is $\xi_{H} = \partial/\partial t$. In the traditional case with a constant $G$, the boundary integral at spatial infinity directly yields the variation of the ADM mass. However, since $G$ is now treated as a variable, both the symplectic potential $\Theta^\mu$ and the Noether charge $Q_\xi^{\mu\nu}$ contain an overall prefactor of $1/G$. Consequently, the total variation can be decomposed into a standard metric variation and an explicit variation with respect to $G$.

Substituting Eqs.\eqref{einsteintheta} and \eqref{einq} into Eq.\eqref{SAdS} and evaluating the boundary terms at the asymptotic boundary $S_r$, we obtain
\begin{equation}
	\begin{aligned}
	\int_{S_r}\left(\tilde{\delta} \mathbf{Q}_{\xi_H}- {\xi_H}\cdot \boldsymbol{\Theta}[\tilde{\delta} \phi] \right)&=\frac{1}{G}(G\tilde{\delta}M + M\tilde{\delta}G)-\frac{1}{G}(\frac{M}{2} + \frac{r^3}{2Gl^2})\tilde{\delta}G \\
	&=\tilde{\delta}M+\frac{M}{2G}\tilde{\delta}G-\frac{r^3}{2 G^2 l^2}\tilde{\delta} G,\label{Sr}
	\end{aligned}
\end{equation}
where the basic thermodynamic relation $\int_{S_H} \tilde{\delta} \mathbf{Q}_{\xi_H} = T\tilde{\delta} S$ has been rigorously proven within the Wald entropy framework. Furthermore, evaluating the partial derivative of the SAdS Lagrangian with respect to Newton's constant $G$, denoted by $F_g \equiv (\partial \mathbf{L}/\partial G)_{g_{\mu\nu}}$, and integrating this quantity over the geometric volume from the black hole horizon $r_h$ to the spatial cutoff radius $r$, we find
\begin{align} 
	-\tilde{\delta}G \int_{\Sigma}F_g{\xi_H}\cdot\boldsymbol{\epsilon}=-\tilde{\delta} G(\frac{3}{8\pi G^2 l^2} \times 4\pi \times \frac{r^3 - r_h^3}{3})=-\frac{r^3}{2 G^2 l^2}\tilde{\delta} G+\frac{r_h^3}{2 G^2 l^2}\tilde{\delta} G.\label{G}
\end{align}
After simplifying and reorganizing Eqs.\eqref{Sr} and \eqref{G}, we arrive at
\begin{align}
        \tilde{\delta} M=T\tilde{\delta} S+\left(-\frac{M}{2G}+\frac{r_h^3}{2 G^2 l^2}\right)\tilde{\delta} G. \label{delt}
\end{align}
To further analyze the thermodynamic relations of the black hole system, we invoke the holographic dictionary, which establishes the relation $C \propto l^2/G$ for the central charge of the boundary field theory \cite{Myers:2010xs,Myers:2010tj}. The detailed theoretical construction is provided in Appendix \ref{holographic principle}. In the restricted phase space of black hole thermodynamics, the quantity $C$ is treated as a new thermodynamic variable. Then, substituting $M$ from Eq.\eqref{M} into Eq.\eqref{delt}, we obtain
\begin{align} 
\tilde{\delta}M=T\tilde{\delta}S+\left(\frac{r_h}{4 l^2}-\frac{r_h^3}{4 l^4}\right)\tilde{\delta}C=T\tilde{\delta}S+\mu_g\tilde{\delta}C,
\end{align}
where $\mu_g$ is identified as the chemical potential conjugate to the central charge $C$.     

On the other hand, the chemical potential can be independently evaluated from the on-shell Euclidean action of the bulk spacetime via the holographic relation $\mu C = T I_E$. The regularized Euclidean action for the four-dimensional Schwarzschild-AdS black hole is calculated by evaluating the bulk, Gibbons-Hawking-York boundary, and counterterm contributions \cite{york1986black}. Based on this rigorous holographic derivation, the chemical potential $\mu$ is calculated as follows, for more details, please refer to Appendix \ref{SAdS:euclidean_action}.
\begin{equation}
        \mu = \frac{T I_E}{C} = \frac{r_h}{4l^2} \left( 1 - \frac{r_h^2}{l^2} \right).
\end{equation}
It is straightforward to verify that this result perfectly matches the chemical potential $\mu_g$ obtained earlier through the extended Iyer-Wald formalism, where
\begin{equation}
        \mu_g = \frac{r_h}{4 l^2} - \frac{r_h^3}{4 l^4}.
\end{equation}
This precise consistency strictly validates the geometric origin of the chemical potential in the thermodynamics of the restricted phase space. Additionally, from Eq.\eqref{Euler}, we can obtain the homogeneous Euler relation corresponding to traditional thermodynamics
\begin{align}
        M=TS+\mu_g C.
\end{align}
Thus, for the fundamental SAdS black hole, we have derived the first law of thermodynamics and the Euler relation directly from the first principles of gravity, without introducing any ad hoc thermodynamic assumptions, thereby establishing a geometric framework for extensive black hole thermodynamics.

\subsection{Reissner-Nordström-AdS black hole}
We next consider the RN-AdS black hole, which introduces a $U(1)$ gauge field on top of the gravitational background. This allows us to verify whether the additional Noether charge from the matter sector can be naturally incorporated into the extended geometric framework without breaking thermodynamic consistency.

The action of 4-dimensional RN-AdS black holes is expressed as
\begin{align}
	I=\frac{1}{16\pi G}\int_{\mathcal{M}}d^4x \sqrt{-g}(R-2\Lambda)-\frac{1}{16 \pi}\int_{\mathcal{M}}d^4x \sqrt{-g}F_{\mu\nu}F^{\mu\nu},
\end{align}
where $F_{\mu \nu}= \partial_{\mu}A_{\nu}-\partial_{\nu}A_{\mu}$ is the Maxwell field tensor. The RN-AdS black hole solution is
\begin{align}
	d s^{2}&=-f(r) d t^{2}+\frac{1}{f(r)} d r^{2}+r^{2}(d\theta^{2}+\sin^2{\theta}d\phi^{2}),\\
	A_{\mu}&=(\Phi(r),0,0,0), 
\end{align}
the metric function and the charge potential are
\begin{align}
	f(r)=1-\frac{2GM}{r}+\frac{G Q^2}{r^2}+\frac{r^2}{l^2},~~~~~\Phi(r)=\frac{Q}{r}.
\end{align}
The related thermodynamic quantities are given by
\begin{align}
	M = \frac{r_h}{2G}\left(1 + \frac{GQ^2}{r_h^2} + \frac{r_h^2}{l^2}\right),\quad
	T = \frac{1}{4\pi r_h}\left(1 - \frac{GQ^2}{r_h^2} + \frac{3r_h^2}{l^2}\right), \quad S = \frac{\pi r_h^2}{G}.
\end{align}
Similarly, we start from the variation $\delta (\sqrt{-g} L)$. Unlike the SAdS black hole, the electromagnetic field introduces additional contributions to the boundary terms. Consequently, the total symplectic potential depends not only on the metric variation $\delta g_{\alpha\beta}$ but also incorporates the variation of the gauge field $\delta A_\nu$
\begin{align}
	\Theta^\mu(\delta g_{\alpha\beta}, \delta A_\nu) \equiv \frac{1}{16\pi G} \left( g^{\mu\alpha} \nabla^\beta \delta g_{\alpha\beta} - g^{\alpha\beta} \nabla^\mu \delta g_{\alpha\beta} \right) - \frac{1}{4\pi G} F^{\mu\nu} \delta A_\nu. \label{RNtheta}
\end{align}
Under a diffeomorphism generated by $\xi$, the variations of the metric and gauge field are given by $\delta g_{\alpha\beta} = \mathcal{L}_\xi g_{\alpha\beta}$ and $\delta A_\nu = \mathcal{L}_\xi A_\nu$. Using Cartan's formula, the Lie derivative of the gauge field can be decomposed into a physical response term and a pure gradient term corresponding to a local $U(1)$ gauge transformation, as shown below
\begin{align}
\mathcal{L}_\xi A_\nu = \xi^\rho F_{\rho\nu} + \nabla_\nu (\xi^\rho A_\rho).
\end{align}
Substituting this expansion into the electromagnetic sector of the symplectic potential and applying the Leibniz rule, the pure gradient contribution is reorganized into an antisymmetric total derivative and a divergence term
\begin{align}
- \frac{1}{4\pi G} F^{\mu\nu} \mathcal{L}_\xi A_\nu = - \frac{1}{4\pi G} F^{\mu\nu} (\xi^\rho F_{\rho\nu}) + \nabla_\nu \left[ - \frac{1}{4\pi G} (\xi^\rho A_\rho) F^{\mu\nu} \right] + \frac{1}{4\pi G} (\xi^\rho A_\rho) \nabla_\nu F^{\mu\nu}.
\end{align}
Constructing the Noether current via $J_\xi^\mu = \Theta^\mu - \xi^\mu L$, the bulk contributions evaluate to combinations of the equations of motion. Under the on-shell condition, both the gravitational field equations and the Maxwell equations ($\nabla_\nu F^{\mu\nu}=0$) vanish. Consequently, the Noether current reduces to a pure divergence form $J_\xi^\mu = \nabla_\nu Q_\xi^{\mu\nu}$, yielding the generalized macroscopic Noether charge \cite{Hawking:1995ap}
\begin{align}
	Q_\xi^{\mu\nu} \equiv -\frac{1}{16\pi G} (\nabla^\mu \xi^\nu - \nabla^\nu \xi^\mu) - \frac{1}{4\pi G} (\xi^\rho A_\rho) F^{\mu\nu}.
	\end{align}
We substitute the aforementioned symplectic potential $\Theta^\mu$ and Noether charge $Q_\xi^{\mu\nu}$ into Eq.\eqref{SAdS}. As with the uncharged SAdS black hole, the horizon integral yields $-T\tilde{\delta}S$ via the Wald entropy formulation, since $\xi_H=0$ at the horizon. At the spatial boundary $S_r$, integrating the variations of the gravitational Noether charge and the symplectic potential produces the mass variation alongside terms dependent on the variation of $G$. Additionally, the electromagnetic sector of the RN-AdS black hole introduces an extra variational contribution at $S_r$, evaluating directly to $-\Phi(r)\tilde{\delta}Q = -\frac{Q}{r}\tilde{\delta}Q$. Thus, it can be concluded that
\begin{align}
		\int_{S_r}\left(\tilde{\delta} \mathbf{Q}_{\xi_H} - {\xi_H}\cdot \boldsymbol{\Theta}[\tilde{\delta} \phi] \right) - \int_{S_H} \tilde{\delta} \mathbf{Q}_{\xi_H}&=\tilde{\delta} M+\frac{M}{2G}\tilde{\delta}G-\frac{r^3}{2 G^2 l^2}\tilde{\delta} G-\frac{Q}{r}\tilde{\delta} Q -T\tilde{\delta} S.   
	\end{align}
Similarly, as calculated by Eq.\eqref{G}, we obtain the volume component due to the variation of $G$
\begin{align}
	-\tilde{\delta}G \int_{\Sigma}F_g{\xi_H}\cdot\boldsymbol{\epsilon}=-&\frac{r^3}{2 G^2 l^2}\tilde{\delta} G+\frac{r_h^3}{2 G^2 l^2}\tilde{\delta} G,
\end{align}
here, $F_g$ is the partial derivative of the total Lagrangian with respect to $G$. Here we introduce the rescaled potential and charge\cite{Cong:2021fnf}, defined as $\hat{\Phi}=\frac{\Phi(r_h) \sqrt{G}}{l}$ and $\hat{Q}=\frac{Q l}{\sqrt{G}}$ \footnote{To maintain consistency with the holographic dictionary, we adopt the same rescaled definitions for the potential and charge as in Ref.~\cite{Cong:2021fnf}.}. By substituting the relation $Q = \hat{Q} \sqrt{G}/{l}$ into the above equation, we can obtain
\begin{equation}
		\begin{aligned}
			\tilde{\delta}M&=T\tilde{\delta}S+\frac{Q}{r}\tilde{\delta}\left(\frac{\hat{Q}\sqrt{G}}{l}\right)+\left(\frac{r_h^3}{4 G^2 l^2}-\frac{r_h}{4 G^2}-\frac{Q^2}{4Gr_h}\right)\tilde{\delta} G\\
			&=T\tilde{\delta}S+\hat{\Phi}\tilde{\delta}\hat{Q}+\left(\frac{r_h^3}{4 G^2 l^2}-\frac{r_h}{4 G^2}+\frac{Q^2}{4Gr_h}\right)\tilde{\delta} G.
		\end{aligned}
\end{equation}
Furthermore, by using the relationship between the central charge and  Newton's constant ($G = l^2/C$) and conducting further simplifications, we arrive at the following expression
\begin{align}
	\tilde{\delta}M=T\tilde{\delta}S+\hat{\Phi}\tilde{\delta}\hat{Q}+\left(\frac{r_h}{4 l^2}-\frac{r_h^3}{4 l^4}-\frac{G Q^2}{4 r_h l^2}\right)\tilde{\delta}C=T\tilde{\delta}S + \hat{\Phi}\tilde{\delta}\hat{Q} + \mu_g\tilde{\delta}C,
\end{align}
where the chemical potential $\mu$ for the RN-AdS black hole can be evaluated from the on-shell Euclidean action, with the detailed regularization process provided in Appendix \ref{RNAdS:euclidean_action}. 
\begin{equation}
	\mu = \frac{T I_E}{C} = \frac{r_h}{4 l^2} \left( 1 - \frac{r_h^2}{l^2} - \frac{G Q^2}{r_h^2} \right).
\end{equation}
By comparison, we find that this precisely matches the chemical potential $\mu_g$ derived via the extended Iyer-Wald formalism. The correctness of obtaining thermodynamic quantities from geometric methods was verified.

Subsequently, from Eq.\eqref{Euler}, the corresponding homogeneous Euler relation is obtained
\begin{align}
	M=TS+\hat{\Phi} \hat{Q}+\mu_g C.
\end{align}
Therefore, even after incorporating the $U(1)$ gauge field, our generalized geometric framework naturally incorporates the associated Maxwell boundary terms and Noether charges.

\subsection{Kerr-AdS black hole}
The Kerr-AdS black hole is a stationary solution of Einstein gravity and has a distinct geometric structure, making it worthy of in-depth study. Its metric can be expressed as follows
\begin{equation}
		\begin{aligned}
			d s^2= & -\frac{\Delta_r}{\rho^2}\left(d t-\frac{a \sin ^2 \theta}{\Xi} d \phi\right)^2+\frac{\rho^2}{\Delta_r} d r^2+\frac{\rho^2}{\Delta_\theta} d \theta^2+\frac{\Delta_\theta \sin ^2 \theta}{\rho^2}\left(a d t-\frac{r^2+a^2}{\Xi} d \phi\right)^2,
		\end{aligned}
\end{equation}
each of these items is defined as
	\begin{align}
			\rho^2&=r^2+a^2 \cos ^2 \theta,~~~~~\Delta_r=\left(r^2+a^2\right)\left(1+\frac{r^2}{l^2}\right)-2 G m r, \\
		\Xi&=1-\frac{a^2}{l^2}, ~~~~~~~~~~~~~~\Delta_\theta=1-\frac{a^2}{l^2} \cos ^2 \theta,
	\end{align}
where $a$ is the rotation parameter of the black hole. The mass $M$ and angular momentum $J$ are related to the parameters $m, a$ via the following expressions \cite{Gibbons:2004ai}
\begin{align}
M=\frac{m}{\Xi^2},~~~~~J=	\frac{a m}{\Xi^2},  \label{mj}
\end{align}
in which $m$ can be solved from the equation $\Delta_r(r_h) = 0$. Therefore, we can write the $M$ and $J$ in terms of $a,\;r_h,\;G$ as 
\begin{eqnarray}
	&& M=\frac{1}{2\Xi^2 G r_h}\left(r_h^2+a^2+\frac{r_h^{4}}{\ell^{2}}+\frac{a^2r_h^{2}}{\ell^{2}}\right),\; \\ && J=\frac{a}{2\Xi^2 G r_h}\left(r_h^2+a^2+\frac{r_h^{4}}{\ell^{2}}+\frac{a^2r_h^{2}}{\ell^{2}}\right).
\end{eqnarray}
The rest of the thermodynamic quantities can be written as
\begin{eqnarray}
	\label{therm1}
	&&T=\frac{\kappa}{2 \pi} =\frac{r_h}{4\pi(r_h^2+a^2)}\left(1+\frac{a^2}{\ell^2}+\frac{3r_h^2}{\ell^2}-\frac{a^2}{r_h^2}\right)\, , \\
	&&S = \frac{A}{4G}=\frac{\pi(r_h^2+a^2)}{G\Xi} \, , \quad
	\Omega = \frac{a\Xi}{r_h^2+a^2}+\frac{a}{\ell^2}.
\end{eqnarray}
In Kerr-AdS spacetime, the horizon Killing vector can be decomposed as: $\xi_{H}=\xi_{t}+\Omega \xi_{\phi}$, where $\xi_{t}\equiv\frac{\partial}{\partial t}+\Omega_{\infty}\frac{\partial}{\partial\phi}$ is the modified time-like Killing vector that accounts for the angular velocity at infinity, $\xi_{\phi}\equiv\frac{\partial}{\partial\phi}$ is the rotational Killing vector, and $\Omega\equiv\Omega_{H}-\Omega_{\infty}$ represents the angular velocity measured relative to infinity. 

This decomposition of the horizon Killing vector makes it straightforward to see that
\begin{align}
	\textbf{Q}_{\xi_H} = \textbf{Q}_{\xi_t}+\textbf{Q}_{\xi_\phi}.
\end{align}
Likewise, based on the definition of the Noether current, we extract its divergence term and simplify it into an antisymmetric derivative form, from which the Noether charge is directly read off and substituted into Eq.\eqref{SAdS}, which yields
\begin{equation}
	\begin{aligned}
		&\int^{(reg.)}_{S_\infty}(\tilde{\delta}\textbf{Q}_{\xi_H}-\xi_H\cdot {\bf\Theta}[\tilde{\delta} \phi])-T\tilde{\delta}S\\
		&=\int^{(reg.)}_{S_\infty}(\tilde{\delta}\textbf{Q}_{\xi_t}-\xi_t\cdot {\bf\Theta}[\tilde{\delta} \phi])+\Omega\int^{(reg.)}_{S_\infty}(\tilde{\delta}\textbf{Q}_{\xi_\phi}-\xi_\phi\cdot {\bf\Theta}[\tilde{\delta} \phi])-T\tilde{\delta}S\\
		&=-\tilde{\delta}G\int_{\mathcal{V}_r}F_g {\xi_H}\cdot\boldsymbol{\epsilon}.
	\end{aligned}
\end{equation}
In addition, by observing that the vector $\xi_\phi$ is tangential to the integrating surface, we can evaluate the expression $ \int^{(reg.)}_{S_\infty}  \left(  \delta \mathbf{Q}_{\xi_\phi} -  \xi_\phi \cdot \boldsymbol{\Theta}[\delta \phi] \right)=\delta\,\int^{(reg.)}_{S_\infty}\mathbf{Q}_{\xi_\phi}= -\delta J$, where $J=\frac{m a}{\Xi^2}$. From this, by further calculation, we can construct
\begin{equation}
	\begin{aligned}
	\frac{l^{4}}{(a^{2}-l^{2})^{2}}\tilde{\delta}m
-\frac{4al^{4}m}{(a^{2}-l^{2})^{3}}\tilde{\delta}a
-\frac{l^{2}m}{2G(a^{2}-l^{2})}\tilde{\delta}G
-\Omega \tilde{\delta}J
-T\tilde{\delta}S
=
\frac{r_h(a^{2}+r_h^{2})}
{2G^{2}(l^{2}-a^{2})}
\tilde{\delta}G .
	\end{aligned}
\end{equation}
By considering the relations Eq.\eqref{mj}, we find that
\begin{equation}
	\tilde{\delta}M=\frac{l^4}{(a^2-l^2)^2}\tilde{\delta}m-\frac{4al^4m}{(a^2-l^2)^3}\tilde{\delta}a.
\end{equation}
Furthermore, by utilizing the relationship between the central charge $C$ and Newton's constant $G$($G =l^2/C$), the above equation can be further simplified to
\begin{align}
	\tilde{\delta}M=T\tilde{\delta}S+\Omega \tilde{\delta}J+\frac{(l^2-r_h^2)(a^2+r_h^2)}{4l^2 r_h(l^2-a^2)}\tilde{\delta}C.
\end{align}
Similarly, the chemical potential $\mu$ for the Kerr-AdS black hole can be independently evaluated from the on-shell Euclidean action, with the detailed regularization procedure provided in Appendix \ref{Kerr:euclidean_action}. Utilizing the holographic relation $\mu C = T I_E$ \cite{Gibbons:2004ai}, we obtain
\begin{equation}
	\mu = \frac{(l^2-r_h^2)(a^2+r_h^2)}{4l^2 r_h(l^2-a^2)}.
\end{equation}
By explicitly comparing this with the chemical potential $\mu_g$ derived via the extended Iyer-Wald formalism, we verify the correctness of the obtained result. In addition, utilizing the integrated identity in Eq.\eqref{Euler}, the corresponding homogeneous Euler relation is directly established. Consequently, we can establish the relation
\begin{align}
	\tilde{\delta}M&=T\tilde{\delta}S+\Omega \tilde{\delta}J+\mu_g\tilde{\delta}C,\\
	M&=TS+\Omega J+\mu_g C.
\end{align}
In summary, despite the intricate axisymmetric geometry and rotational dynamics of the Kerr-AdS black hole, the extended Iyer-Wald formalism successfully captures its complete thermodynamic structure. We derived precise expressions for the conjugate chemical potential and rigorously established the homogeneous Euler relation consistent with the first law. These outcomes perfectly align with the restricted phase space results presented in \cite{Gao:2021xtt}.

Based on the above discussion, we have selected three typical black hole models, namely Schwarzschild-AdS, RN-AdS, and Kerr-AdS, which constitute a progressive validation sequence from simple to complex, from uncharged 
to charged, and from static to rotating. Although the introduction of new physical parameters and the progressive breaking of spacetime symmetries lead to different analytical expressions for their respective conjugate chemical potentials $\mu_g$, the underlying geometric logic of the derivations is completely unified. For ease of comparison, Table \ref{tab:summary} collects the explicit forms of $\mu$ for the three cases.
\begin{table}[htbp]
\centering
\caption{A progressive validation of the extended Iyer-Wald formalism.}
\renewcommand{\arraystretch}{2.3}
\setlength{\tabcolsep}{8pt}
\begin{tabular}{c|c|c|c}
\hline
Solution & New ingredient & Chemical potential $\mu$ & Geometric origin \\
\hline
Schwarzschild-AdS & — & $\displaystyle \frac{r_h}{4l^2}\left(1 - \frac{r_h^2}{l^2}\right)$ & Horizon + boundary \\[6pt]
RN-AdS & Charge $Q$ & $\displaystyle \frac{r_h}{4l^2}\left(1 - \frac{r_h^2}{l^2} - \frac{GQ^2}{r_h^2}\right)$ & + U(1) Noether charge \\[6pt]
Kerr-AdS & Rotation $a$ & $\displaystyle \frac{(l^2-r_h^2)(a^2+r_h^2)}{4l^2 r_h(l^2-a^2)}$ & Rotating $\xi_H$ \\[6pt]
\hline
\multicolumn{4}{c}{Same three-step procedure applies to all cases, despite increasing complexity.} \\
\hline
\end{tabular}
\label{tab:summary}
\end{table}

The results show that, despite the different forms of $\mu$ arising from the different spacetime symmetries and matter content, all three cases are governed by the same geometric procedure, demonstrating the universality of the extended Iyer-Wald formalism.

\section{Conclusion}\label{IV}

In this work, we have extended the Iyer-Wald formalism and constructed a self-consistent framework for homogeneous black hole thermodynamics in any diffeomorphism-invariant gravitational theory. Rather than assuming the laws of thermodynamics a priori or phenomenologically deducing them from thermodynamic identities, this framework derives the explicit expressions of all thermodynamic variables directly from geometric principles (the Noether currents and charges arising from the Lagrangian density).

Specifically, we focus on the restricted phase space where the cosmological constant (or the AdS radius $l$) is fixed while Newton’s constant $G$ varies. The Iyer-Wald formalism is extended by systematically introducing the variation of $G$ into the Lagrangian flow and the geometric structure of the Noether charges. This extension enables us to rigorously establish the first law of thermodynamics and the homogeneous Euler relation, thereby ensuring the extensivity that is necessary for alignment with classical thermodynamics.

We demonstrate the consistency of this framework through applications to three canonical black hole solutions, i.e., Schwarzschild-AdS, Reissner-Nordström-AdS, and Kerr-AdS. These three cases form a progressive validation sequence from simple to complex, from uncharged to charged, and from static to rotating. Despite the different explicit forms of the chemical potential in each case, the underlying geometric derivation is unified. For each solution, we derive explicit thermodynamic equations and expressions for the relevant variables. Crucially, we independently compute the chemical potential using the gravitational Euclidean action on the shell and find perfect agreement with the geometrically derived values.

This work is conceptually distinct from previous Iyer-Wald extensions that operate within the extended phase space with $\Lambda$ varying or with $G$ fixed. By focusing on the restricted phase space where $G$ varies and $\Lambda$ is fixed, we provide the missing geometric foundation for a thermodynamically extensive description of black holes. In summary, the successful application of the extended Iyer-Wald formalism highlights its utility as a reliable geometric tool for investigating black hole thermodynamics and provides a rigorous first-principles derivation of black hole thermodynamics with variable Newton's constant. 

Furthermore, it is also worth noting that the roles of the $a$- and $c$- charges in exact black hole thermodynamics remain incompletely understood, particularly in situations where these two charges become decoupled \cite{Li:2018drw}. The geometric approach developed in this work may offer a promising avenue for clarifying their thermodynamic significance. We anticipate that further applications of this approach will help reveal richer black hole thermodynamic properties and provide deeper insights into their underlying microscopic structures.

\section*{Acknowledgement}
\vspace{-0.7em}
We are grateful to the anonymous reviewers and editor for their constructive suggestions, which have greatly improved our manuscript. This research was supported in part by the National Natural Science Foundation of China (Grant Nos.12275216, 12575064, 12247103), by the Natural Science Basic Research Plan in Shaanxi Province of China (Grant No. 2025JC-YBQN-029).
\end{spacing}

\begin{appendices}
	\section{Holographic Interpretation of the Chemical Potential}\label{holographic principle}
	The holographic dictionary establishes a profound structural isomorphism, dictating a strict equivalence between the thermal partition function of the boundary conformal field theory (CFT) at finite temperature and the on-shell partition function of the bulk gravity theory \cite{Gubser:1998bc,Witten:1998qj}, namely
	\begin{equation}
		Z_{\rm CFT} = Z_{\rm AdS}.
	\end{equation}
	Based on this core equivalence relation, we can further construct the connection between macroscopic thermodynamic quantities and bulk geometric quantities. In the boundary CFT, the macroscopic thermodynamic free energy $W$ of the system can be defined by the standard statistical physics partition function as 
	\begin{equation}
		W = -T \ln Z_{\rm CFT}.
	\end{equation}
	On the gravity side, under the semi-classical approximation, the gravitational path integral is dominated by the contribution of the classical saddle point. Therefore, the partition function of the bulk spacetime is completely determined by the on-shell Euclidean action $I_E$, yielding
	\begin{equation}
		I_E = -\ln Z_{\rm AdS}.
	\end{equation}
	By combining the two, we naturally establish the connection between the free energy and the gravitational action, given by
		\begin{equation}
			W = -T \ln Z_{CFT} = -T \ln Z_{AdS} = T I_E.
		\end{equation}
		
		On this basis, when introducing the AdS radius $l$ as the length scale, for a $d$-dimensional AdS spacetime, the particle degrees of freedom in the boundary field theory can be deduced as $N^2 \propto l^{d-2}/G$. In conformal field theory, the central charge $C$ essentially measures the microscopic degrees of freedom of the system, thus establishing the scaling relation between the central charge and the bulk geometric parameters as \cite{Henningson:1998gx,Freedman:1999gp,Myers:2010xs,Myers:2010tj}
		\begin{equation}
			C \propto l^{d-2}/G,
		\end{equation}
		where the proportionality coefficient depends on the specific holographic system and its normalization convention. For the four-dimensional gravity system focused on in this paper (whose holographic boundary is three-dimensional, i.e., $d=3$), the scaling relation is $C \propto l^2/G$.
		
		Building upon this foundation, if the central charge $C$ is treated as an independent thermodynamic variable, the conjugate thermodynamic quantity is the chemical potential $\mu$, and the combination of these two constitutes the free energy $W=\mu C$ of the system\cite{Karch:2015rpa}. We can obtain the following relation
	\begin{equation}
		\mu C = W = T I_E.
	\end{equation}
	Therefore, by calculating the on-shell Euclidean action of the bulk gravity, the chemical potential $\mu$ can be directly and self-consistently extracted from this Euclidean relation, and the corresponding Smarr relation is subsequently obtained to verify the correctness of the results derived under our geometric framework.
	
	\section{Derivation of the Euclidean Action}\label{app:euclidean_action}
	
	\subsection{The Complete Euclidean Action and Boundary Terms}
	\setcounter{equation}{0}
	We start by performing a Wick rotation $t \to i\tau$ to obtain the Euclidean metric, where the Euclidean time is periodic with period $\beta = 1/T$ to avoid conical singularities at the event horizon. The complete Euclidean action encompasses the Einstein-Hilbert bulk term $I_{\rm bulk}$, the Gibbons-Hawking-York boundary term $I_{\rm GHY}$, the Maxwell electromagnetic bulk term $I_{\rm EM}$, the Maxwell boundary term $I_{\rm EM,bdy}$ and the counterterm $I_{\rm ct}$ required for AdS regularization
	\begin{equation}
		\begin{aligned}
			I_E &= I_{\rm bulk}+ I_{\rm GHY} + I_{\rm EM} +I_{\rm EM,bdy}+ I_{\rm ct} \\
			&= - \frac{1}{16\pi G} \int_{\mathcal{M}} d^D x \sqrt{g} (R - 2\Lambda) - \frac{1}{16\pi G} \int_{\mathcal{M}} d^D x \sqrt{g} F_{\mu\nu} F^{\mu\nu}\\
			&\quad - \frac{1}{8\pi G} \int_{\partial \mathcal{M}} d^{D-1} x \sqrt{h} K - \frac{1}{8\pi G} \int_{\partial \mathcal{M}} d^{D-1} x \sqrt{h} n_\mu F^{\mu\nu} A_\nu \\
			&\quad + \frac{1}{8\pi G} \int_{\partial \mathcal{M}} d^{D-1}x \sqrt{h} \left( \frac{D-2}{l} + \frac{l}{2(D-3)} \mathcal{R} + \mathcal{O}(\mathcal{R}^2) \right),
		\end{aligned}
	\end{equation}
	where $D$ denotes the spacetime dimension, $F_{\mu\nu} = \partial_\mu A_\nu - \partial_\nu A_\mu$ is the electromagnetic field strength tensor, $n^\mu$ is the unit normal vector to the boundary. $h$ is the determinant of the induced metric $h_{ab}$ on $\partial \mathcal{M}$, and $K$ is the trace of the extrinsic curvature. The quantity $\mathcal{R}$ represents the Ricci scalar constructed from the induced metric $h_{ab}$.
	
	\subsection{Specific Calculations of the Euclidean Action}
	
	\subsubsection{Schwarzschild-AdS (SAdS) Black Hole}\label{SAdS:euclidean_action}
	For the four-dimensional SAdS black hole, by evaluating the bulk and the Gibbons-Hawking-York boundary terms up to a finite spatial cutoff $r = r_c$, we obtain\cite{york1986black} 
	\begin{equation}
			\begin{aligned}
				I_{\rm bulk} + I_{\rm GHY}&=- \frac{1}{16\pi G} \int_{\mathcal{M}} d^4 x \sqrt{g} (R - 2\Lambda) - \frac{1}{8\pi G} \int_{\partial \mathcal{M}} d^3 x \sqrt{h} K\\
				&=\frac{\beta}{2G} \left( - \frac{2r_c^3}{l^2} - 2r_c + 3GM - \frac{r_h^3}{l^2} \right),
			\end{aligned}
	\end{equation}
	where $\beta$ is the inverse Hawking temperature and $r_h$ is the horizon radius. As $r_c \to \infty$, this unregularized action exhibits expected spatial divergences. To eliminate these divergences, we evaluate the holographic counterterm $I_{\rm ct}$ on the boundary $r=r_c$ and expand it asymptotically
	\begin{align}
		I_{\rm ct} = \frac{1}{8\pi G} \int_{\partial \mathcal{M}} d^3x \sqrt{h} \left( \frac{2}{l} + \frac{l}{2} \mathcal{R} + \mathcal{O}(\mathcal{R}^2) \right)=\frac{\beta}{2G} \left( \frac{2r_c^3}{l^2} + 2r_c - 2GM + \mathcal{O}(r_c^{-1}) \right).
	\end{align}
	By summing these three contributions, the divergent terms scaling with $r_c$ cancel exactly. Substituting into the expressions of the relevant thermodynamic quantities and taking the limit $r_c \to \infty$, we arrive at the finite regularized Euclidean action
	\begin{equation}
		I_E = \lim_{r_c \to \infty} (I_{\rm bulk} + I_{\rm GHY} + I_{\rm ct}) = \frac{\beta}{2G} \left( GM - \frac{r_h^3}{l^2} \right) = \frac{\beta r_h}{4G} \left( 1 - \frac{r_h^2}{l^2} \right).
	\end{equation}
	According to the established holographic relation, the chemical potential $\mu$ conjugate to the central charge $C$ can be extracted directly via $\mu C = T I_E$. Substituting the inverse Hawking temperature $T = \beta^{-1}$ and the central charge definition $C = l^2/G$, we explicitly obtain
	\begin{equation}
		\mu = \frac{T I_E}{C} = \frac{r_h}{4l^2} \left( 1 - \frac{r_h^2}{l^2} \right).
	\end{equation}
	By comparing the chemical potential obtained through the extended Iyer–Wald formalism method with the actual one, we can verify the correctness of our method.
	
	\subsubsection{Reissner-Nordström-AdS (RN-AdS) Black Hole}\label{RNAdS:euclidean_action}
	For the four-dimensional charged RN-AdS black hole, following the identical regularization procedure, the total action is evaluated at the cutoff $r = r_c$. Taking the limit $r_c \to \infty$ ensures that all divergent terms precisely cancel. By utilizing the event horizon condition $f(r_h) = 0$ to eliminate the mass parameter $M$, the relation evaluates to \cite{Chamblin:1999tk}
	\begin{equation}
		\begin{aligned}
			T I_E &= \frac{1}{\beta} \lim_{r_c \to \infty} (I_{\rm bulk} +  I_{\rm GHY} +I_{\rm EM}+I_{\rm EM,bdy} + I_{\rm ct}) \\
			&= \frac{1}{\beta} \lim_{r_c \to \infty} \left\{ \frac{\beta}{2G} \left[ \underbrace{\left( - \frac{2r_c^3}{l^2} - 2r_c + 3 G M - \frac{r_h^3}{l^2} - \frac{G Q^2}{r_h} \right)}_{\text{Bulk + EM + GHY}} + \underbrace{\left( \frac{2r_c^3}{l^2} + 2r_c - 2 G M \right)}_{\text{Counterterm } I_{\rm ct}} \right] \right\} \\
			&= \frac{1}{2G} \left( G M - \frac{r_h^3}{l^2} - \frac{G Q^2}{r_h} \right) = \frac{r_h}{4 G} \left( 1 - \frac{r_h^2}{l^2} - \frac{G Q^2}{r_h^2} \right),
		\end{aligned}
	\end{equation}
	where $Q$ represents the electric charge. Using $\mu C = T I_E$, we obtain
	\begin{equation}
		\mu = \frac{T I_E}{C} = \frac{r_h}{4 l^2} \left( 1 - \frac{r_h^2}{l^2} - \frac{G Q^2}{r_h^2} \right).
	\end{equation}
	This once again demonstrates that by comparing the chemical potential $\mu$ calculated from the Euclidean action with the chemical potential $\mu_g$ obtained using the extended Iyer-Wald geometric form method, the validity of the results can be further verified.
	
	\subsubsection{Kerr-AdS Black Hole}\label{Kerr:euclidean_action}
	The Kerr-AdS black hole is a stationary solution of Einstein gravity characterized by a distinct axisymmetric geometric structure. Following the same regularization procedure, the total unregularized action is evaluated at the spatial cutoff $r=r_c$. Taking the limit $r_c \to \infty$, we obtain\cite{Gibbons:2004ai}
	\begin{equation}
		\begin{aligned}
			T I_E &= \frac{1}{\beta} \lim_{r_c \to \infty} (I_{\rm bulk} + I_{\rm GHY} + I_{\rm ct}) \\
			&= \frac{1}{\beta} \lim_{r_c \to \infty} \left\{ \frac{\beta}{2G\Xi} \left[ \underbrace{\left( - \frac{2r_c^3}{l^2} - 2r_c + 3 G m - \frac{r_h(r_h^2+a^2)}{l^2} \right)}_{\text{Bulk + GHY}} + \underbrace{\left( \frac{2r_c^3}{l^2} + 2r_c - 2 G m \right)}_{\text{Counterterm } I_{\rm ct}} \right] \right\} \\
			&= \frac{1}{2G\Xi} \left( 3 G m - \frac{r_h(r_h^2+a^2)}{l^2} - 2 G m \right) \\
			&= \frac{1}{2G(1 - a^2/l^2)} \left[ \frac{(r_h^2+a^2)(l^2+r_h^2)}{2 r_h l^2} - \frac{2r_h^2(r_h^2+a^2)}{2r_h l^2} \right] = \frac{(l^2-r_h^2)(a^2+r_h^2)}{4 G r_h (l^2-a^2)},
		\end{aligned}
	\end{equation}
	where $\Xi=1-\frac{a^2}{l^2}$ and $a$ represents the rotation parameter. The mass $M$ is related to the parameters $m$ and $a$ via $M=\frac{m}{\Xi^2}$. We observe that all divergent terms cancel out precisely, contributing to the relevant finite thermodynamic quantities of the Kerr-AdS black hole. We thereby obtain the simplified chemical potential $\mu$
	\begin{equation}
		\mu = \frac{T I_E}{C} = \frac{(l^2-r_h^2)(a^2+r_h^2)}{4 l^2 r_h(a^2-l^2)}.
	\end{equation}
		\begin{equation}
		\mu = \frac{T I_E}{C} = \frac{(l^2-r_h^2)(a^2+r_h^2)}{4 l^2 r_h(l^2-a^2)}.
	\end{equation}
	Similarly, by comparing this with the chemical potential $\mu_g$ derived from the extended Iyer-Wald geometric formalism for the Kerr-AdS solution, we can verify the correctness of our result.
\end{appendices}

\end{document}